\documentclass[runningheads]{llncs}
\usepackage{graphicx}
\usepackage{xcolor}
\usepackage{multirow}
\usepackage{comment}
\usepackage[T1]{fontenc}
\usepackage{lmodern}
\usepackage{textcomp}
\usepackage{bookmark}
\usepackage{orcidlink}
\usepackage[backend=biber,style=numeric,sorting=none]{biblatex}
\bibliography{main}

\begin{document}
\title{QEQR: An Exploration of Query Expansion Methods for Question Retrieval in CQA Services}
\titlerunning{Question Expansion}  
\author{Yasin Ghafourian\inst{1}\orcidlink{0000-0001-9683-9748} \and Sajad Movahedi\inst{1} \and
Azadeh Shakery\inst{1}\orcidlink{0000-0003-1799-8340}}

\authorrunning{Ghafourian et al.} 


\institute{University of Tehran, Tehran, Iran,\\
\email{\{yasin.ghafouryan,s.movahedi,shakery\}@ut.ac.ir}}
\maketitle              
\begin{abstract}
CQA services are valuable sources of knowledge that can be used to find answers to users' information needs. In these services, question retrieval aims to help users with their information need by finding similar questions to theirs. However, finding similar questions is obstructed by the lexical gap that exists between relevant questions. In this work, we target this problem by using query expansion methods. We use word-similarity based methods, propose a question-similarity based method and selective expansion of these methods to expand a question that's been submitted and mitigate the lexical gap problem. Our best method achieves significant relative improvement of 1.8\% compared to the best performing baseline without query expansion.
\keywords {Community Question Answering, Question Retrieval, Question Expansion, Word Embedding}
\end{abstract}
%

\section{Introduction}
In the past few years, online services where users are able to contribute in the development of online contents have gained popularity. Community-based Q\&A (CQA) services (e.g. Yahoo! Answers\footnote{\url{http://www.answers.yahoo.com}}) are one of those web services that help millions of users with their information needs. Users can either post questions in these websites or help solve another user's question.
\par The ongoing use of these CQA services have accumulated a large archive of questions and their corresponding answers that can be effectively used to answer similar questions in the future. Question retrieval in these systems is a means by which users can utilize this valuable archive. Users can use a question retrieval module to see whether their question has already been asked in the system in the form of a similar question. Question retrieval, however, is facing some challenges in its way, one of which is called lexical gap in the literature. Lexical gap happens when two questions with the same intention differ in terms of words used to express that intention \cite{srba2016comprehensive}.

\par To tackle the lexical gap problem, different methods have been proposed over the years. Since the questions are often verbose and there might seem no need to expand them, to the best of our knowledge, there has been no work on question expansion to deal with the lexical gap problem. In this research, we focus on expanding the questions that users submit in order to find more-relevant questions that already exist in the archive and can help users find answers to their information needs faster. Our contributions are four-fold: I) we explore the already proposed query expansion methods to deal with the lexical gap problem, II) we propose a question-similarity based method to expand the questions, III) we use contextualized word embedding for question expansion , and IV) We introduce selective expansion for question expansion methods.

\section{Previous Studies}

\subsection{Question Retrieval}
In the literature different approaches have been adopted to resolve the lexical gap problem that hinders the full performance of the question retrieval system. Some researches consider translation based approaches such as \cite{jeon2005finding,xue2008retrieval,singh2012entity,park2015using}. There are also other works that exploit questions' topics to alleviate the problem, to name a few \cite{cai2011learning,ji2012question,zhang2014topic}. Another batch of works use metadata such as question categories to find similar questions easier, some of which are \cite{cao2009use,cao2012approaches,zhou2013towards,zhou2014group}. Some of the works take into account the syntactic structures of the questions to find questions similar to them like
\cite{duan2008searching,wang2010segmentation}. Finally, there are deep learning based methods that have drawn attention in more recent works \cite{zhou2017modeling,zhang2016learning,wang2017concept,yang2017neural}. 
\subsection{Query Expansion}
In the current information retrieval systems, user queries are mainly processed by using exact matches. This causes a problem of term mismatching, also dubbed as the vocabulary problem \cite{furnas1987vocabulary}.
\par One of the remedies for the vocabulary problem is query expansion, which aims to enrich the query by adding more relevant words to it. The earliest mention of such methods was in the seminal works of Maron and Kuhn \cite{maron1960relevance}, and Rocchio \cite{rocchio1971relevance}.
\par The search engine era introduced another challenge for information retrieval: while the queries were similar in terms of content, the amount of data increase with an exponential rate. This caused the vocabulary problem to become even more severe, given the synonymy and polysemy property of words. Therefore, the precision of most of the previous methods decreased significantly \cite{harman1992relevance}, and the query expansion methods became an even more appealing way of dealing with it 
\cite{azad2019query}. For a comprehensive study of query expansion in information retrieval refer to \cite{azad2019query}.
\subsection{Word and Sentence Representation}
Due to the discrete nature of natural languages, representing the tokens (words, phrases, paragraphs, etc.) in a way that would convey useful syntactic and semantic information, is an active line of research in Natural Language Processing and Information Retrieval \cite{mikolov2013distributed}\cite{peters-etal-2018-deep}\cite{cer-etal-2018-universal}. 
\par The two most common methods of word representation based on co-occurrence of words are word2vec models \cite{mikolov2013distributed} and \cite{pennington-etal-2014-glove}. One of the shortcomings of representing words via these methods is the lack of sentence level features. Due to the context dependent nature of semantic and syntactic role of words, this can potentially become a source of error \cite{peters-etal-2018-deep}. In order to alleviate this problem, it has recently been proposed to use pre-trained neural language models for a context dependent representation of words. It has been empirically proved by \cite{peters-etal-2018-deep}\cite{howard2018universal} and other recent publications introducing new methods of representation based on other variants of neural language models (NLM) \cite{yang2019xlnet}\cite{devlin-etal-2019-bert}, that the latent features extracted by a pre-trained NLM contain rich semantic and syntactic information.

\section{Question Expansion Methods}
In this section we introduce the methods utilized to expand an input question in order to relieve the lexical gap problem. At first, we expand the input question using the most semantically similar words to it that exist in the collection, for which we exploit two of the existing methods put forth by \cite{almasri2016comparison} and \cite{kuzi2016query}. Afterwards, we propose expanding the input question using the words of the most semantically similar questions in the collection to the input question. Finally, we discuss the selective expansion of the terms in the input question following the idea that expanding some of the terms might cause the question's intent to change and that is not favourable in the task of finding questions with similar information needs.\par In all of the proposed methods, an expanded language model for the input question is created and the relevance score between this expanded question and each of the candidate questions in the collection is calculated using the KL-Divergence formula \cite{lafferty2001document,zhai2007notes} with Dirichlet-Prior smoothing as follows:
\begin{equation}
\label{KLDFormula}
Sim\left({\hat{Q}}^{in},\,\,Q^{C} \right) =-D\left( \theta _{{\hat{Q}}^{in}}||\theta _{Q^{C}} \right)
\end{equation}
which becomes:
$$
Sim\left({\hat{Q}}^{in},\,\,Q^{C} \right) =\sum_{w\in Q^{C},\,\,p\left( w|\theta _{{\hat{Q}}^{in}} \right) >0}{\left[ p\left( w|\theta _{{\hat{Q}}^{in}} \right) \log \frac{p_{seen}\left( w|\theta_{Q^{C}} \right)}{\alpha _dp\left( w|C \right)} \right]}+\log \alpha _d
$$
where ${\hat{Q}}^{in}$ is the expanded input question, $\theta _{{\hat{Q}}^{in}}$ is its language model, $Q^{C}$ is a question in the collection and $\theta _{Q^{C}}$ is the language model for that question.
\subsection{Question Expansion Using Similar Words}
The first attempt made in this research to reduce the lexical gap between two relevant questions regards trying to expand the input question with words that are most likely to have been added to the input question by another user with the same information need. We utilize two existing query expansion methods.
\subsubsection{Word-by-word Expansion}
The first method that we utilize to expand questions is suggested by ALmasri et al. \cite{almasri2016comparison} in which each question word, referred to as base word, is expanded individually. For this purpose, $k$ most similar words to each base word are extracted, weighted according to the base word's weight in the question and then added to the expanded question's language model.\par More formally, firstly, a similarity matrix is built that encompasses the similarity between every two words in the collection in terms of their cosine similarity:
\begin{equation}
\label{equ:almasri_SIM}
SIM\left(t_1,t_2\right) = \widetilde{\cos}\left(\overrightarrow{t_1},\overrightarrow{t_2}\right)
\end{equation}
where $\overrightarrow{t_1}$ and $\overrightarrow{t_2}$ are vector representations for two separate words $t_1$ and $t_2$ in the collection and $\widetilde{\cos}$ denotes cosine similarity.
Given a funcion, $top_k(t)$, that returns the $k$ most semantically similar words to the base word, $t$, there will be a set, $top_k\left( t \right) \,\,=\,\,\left[ \acute{t}_1,\,\,\acute{t}_2,\,\,\acute{t}_3,\,\,...\,\,,\,\,\acute{t}_k \right]$, for each base word $t$ that comprises the closest words to $t$ which we refer to as the expansion terms. In order to add these terms to the expanded question language model, firstly, they need a count that also reflects their importance in the language model. To this end, the count of each word is then calculated as follows:
\begin{equation}
\label{equ:almasri_expansioncount}
count\left( \acute{t}_i,\hat{Q}^{in} \right) =\alpha \times count\left( t,Q^{in} \right) \times SIM\left(t_1,\acute{t}_i\right)
\end{equation}
where $\alpha$ is a tuning parameter between $[0,1]$.\par As it can be inferred, the count of each expansion term is mostly influenced by its closeness to the base word and the importance of the base word itself. One issue that requires attention here is that the sum of the counts of all the expansion terms elicited for one base word might exceed the count of the base word and this is not favourable. The reason is that since the base word itself might not be of importance in comparison to other words present in the question, this phenomena shadows the relative importance among terms, through bringing the current base word to a degree of importance close to other words in the question. In this paper, we address this issue by adding a slight modification to formula \ref{equ:almasri_expansioncount} by normalizing the expansion terms' counts so that their sum equals the base word's count:
\begin{equation}
    count^{new}\left( \acute{t}_i,\,\,\hat{Q}^{in} \right) =\frac{count\left( \acute{t}_i,\hat{Q}^{in} \right)}{\sum_{j\in \left[1,k\right]}{count\left( \acute{t}_j,\,\,\hat{Q}^{in} \right)}}\times count\left( t,Q^{in} \right)
\label{equ:correct_AlmasriWeightingscheme} 
\end{equation}
For all base words in the input question, $count^{new}\left(\right)$, is the same as their initial count in the input question:
\begin{equation}
    count^{new}\left( w,\hat{Q}^{in} \right) =\,\,count\left( w,Q^{in} \right)
\label{equ:count_base_words}
\end{equation}
Finally, the expanded language model for the input question is attained via sum-normalizing the counts of all base words and expansion words as follows:
\begin{equation}
p\left( t|\hat{Q}^{in} \right) \,\,=\,\,\frac{count^{new}\left( t,\hat{Q}^{in} \right)}{\sum_{w\in \hat{Q}^{in}}{count^{new}\left( w,\hat{Q}^{in} \right)}}
\label{equ:ALmasri_LM_model}
\end{equation}
We refer to this method as question expansion using ALmasri method (\textbf{expAL}) in our experiments.

\subsubsection{Whole Question Expansion} The second method, proposed by S. Kuzi et al. \cite{kuzi2016query}, uses words that are semantically close to the entire question as opposed to finding close words to each question term individually. In the first step, a vector representation is obtained by length-scale averaging the word representations: 
\begin{equation}
\label{equ:sarkuziCentcalc}
\overrightarrow{{Q}_{Cent}^{in}}  = \sum_{q_i\in Q^{in}} \overrightarrow{{q}_{i}}
\end{equation}

where $\overrightarrow{{q}_{i}}$ is the vector representation for word $q_{i}$ in the input question and $\overrightarrow{{Q}_{Cent}^{in}}$ is the attained vector representation of the input question that reflects the center of the question in the embedding space.\par After that, in order to find the most similar words to the input question, a closeness score between each of the words in the collection and the center of input question is calculated as follows:
\begin{equation}
\label{sarkuzicosinecalc}
S_{Cent}(t;Q^{in}) = \exp(\cos(\overrightarrow{t}, \overrightarrow{{Q}_{Cent}^{in}})) 	
\end{equation}
where $\overrightarrow{t}$ is the L2-normalized Word2Vec vector representing term $t$. Finally, first $v$ words with the highest $S_{Cent}()$ scores are chosen and their scores are sum-normalized to form a probability distribution, $P_{Cent}\left( t|Model \right)$. This probability distribution is then integrated with the language model induced from the input question, $Q^{in}$, using maximum likelihood estimation (MLE), $P_{MLE}\left( t|Q^{in} \right)=\frac{count(t,Q^{in})}{|Q^{in}|}$, to form the language model of the expanded input question:
\begin{equation}
\label{equ:sarkuzifinalmodel}
P\left( t|\hat{Q^{in}}\right)= P\left( t|Q^{in},Model \right) =\lambda P_{MLE}\left( t|Q^{in} \right) +\left( 1-\lambda \right) P_{Cent}\left( t|Model \right)
\end{equation}
We refer to this expansion method in our experiments as question expansion using Kuzi's method (\textbf{expKuzi}).

\subsection{Question Expansion Using Similar Questions}
\subsubsection{Using semantically similar questions}
In our second attempt to reduce the lexical gap between an input question and its relevant questions, we use the questions in the collection that are semantically relevant to the input question for expansion. This method is carried out in four steps: (a)  calculate vector representations of the input question and all the questions in the collection which are expansion candidates; (b) score all the expansion candidates in terms of their cosine similarity to the input question; (c) use the $k$ most similar candidate questions to build a similar-questions' language model; (d) interpolate this language model with the initial language model of the input question. 

\par To achieve a vector representation for each question, contextualized word representations which are based on a pre-trained bi-directional neural language model are used. The reason we did not use the non-contextual word embeddings is the overwhelming evidence of the superiority of the contextual variants \cite{peters-etal-2018-deep}\cite{devlin-etal-2019-bert}\cite{yang2019xlnet}. More specifically, we used ELMo representations \cite{peters-etal-2018-deep} for each word in questions. The reason we used contextualized representations for each question is that there are cases where two questions have the same word but those words have different semantics which is distinguishable according to the contexts in which those words have appeared in \cite{peters-etal-2018-deep} and we need questions with similar contexts to be used for expansion because they are more likely to fill the lexical gap with other words that can be used in the same context. To illustrate, the word "play" in the two queries "Who plays Jacob Black?" and "What instrument does John Mayer play?" convey two completely different meanings. In the former, the word "play" is referring to acting in a movie, while the latter is referring to producing music using a musical instrument. A context independent representation of these two queries will consider the two words "play" similar, which can result in irrelevant candidate questions to the input question getting higher similarity scores for expansion.
\par Considering the input question, $Q^{in}$, and the list of all questions existing in the collection,$\left\{ Q_{1}^{C},\,\,Q_{2}^{C},\,\,Q_{3}^{C},\,\,...\,\,,Q_{M}^{C} \right\}$, we build a contextualized vector representation for each of these questions using the ELMo model\footnote{Implemented by AllenNLP, available at \url{https://github.com/allenai/allennlp/blob/master/tutorials/how_to/elmo.md}.}. For each word in each question, we concatenate the output of each of the LSTM layers at each direction, and use this vector as the representation of the word in the question. We then represent the question by calculating the length-scaled average of its words' contextualized embedding vectors:
\begin{equation}
\label{equ:ELMo_Vector_rep_in}
\overrightarrow{{V}_{ELMo}^{in}}  = \sum_{q_i\in Q^{in}} \overrightarrow{{q}_{i}}
\end{equation}
\begin{equation}
\label{equ:ELMo_Vector_rep_candid}
\overrightarrow{{V}_{ELMo}^{C}}  = \sum_{q_j\in Q^{C}} \overrightarrow{{q}_{j}}
\end{equation}
where $\overrightarrow{{V}_{ELMo}^{in}}$ is a vector representation for the input question, $\overrightarrow{{V}_{ELMo}^{C}}$ is vector representation for a candidate question in the collection, and $q_i$ and $q_j$ are the representation of the $i^{th}$ word of the input question and the $j^{th}$ word of the candidate question, respectively.
\par Now that we have a vector for each question, we rank the entire collection with respect to the input question in terms of the cosine similarity of their vector representations to that of the input question. Consequently, the similarity between each candidate question and the input question is assessed as follows:
\begin{equation}
\label{equ:ELMo_Cosine}
Sim\,\left(Q^{in},Q^{C}\right)= \widetilde{Cos}(\overrightarrow{{V}_{ELMo}^{in}},\overrightarrow{{V}_{ELMo}^{C}})
\end{equation}
The $k$ most similar candidate questions are then segregated to form the set $F_{ELMo} =\,\{Q_{1}^{C},\,\,Q_{2}^{C},\,\,Q_{3}^{C},\,\,Q_{4}^{C},\,\,...\,\,,Q_{k}^{C}\}$ upon which we then build a language model, $\theta_{F_{ELMo}}$, using maximum likelihood estimation. Finally, interpolating this language model with the input question's initial language model gives the expanded language model for the input question (\textbf{expELMo}): 
\begin{equation}
\label{equ:RM3_ELMo}
p\left( t\left| \hat{Q}^{in} \right. \right) =\left( 1-\alpha \right) p\left( t\left| Q^{in} \right. \right) +\alpha p\left( t\left| \theta_{F_{ELMo}} \right. \right)
\end{equation}
\subsubsection{Integration with pseudo relevance feedback} Pseudo relevance feedback has shown to be very effective in enhancing the performance in information retrieval systems \cite{lafferty2001document,lv2009comparative} and  pseudo-feedback-based query expansion methods can be considered complementary to word embedding based query expansion approaches \cite{kuzi2016query}. Therefore, we further extend the input question's language model by incorporating the language model achieved from the pseudo-relevance feedback questions retrieved using a basic information retrieval method. To do so, first we retrieve a number of questions from the collection as the feedback questions' set, $F$, and calculate their language model. Simple mixture model (SMM) \cite{zhai2001model}, is used to calculate the weight of the words in the feedback language model:
\begin{equation}
\label{equ:MixtureModel_base_formula}
\log _p\left( F\left| \theta _F \right. \right) =\sum_{t\epsilon V}{c\left( t,F \right) \log \left( \left( 1-\lambda \right) p\left( t\left| \theta _F \right. \right) +\lambda p\left( t\left| C \right. \right) \right)}
\end{equation}
In this equation, $c\left( t,F \right)$ is the count of word $w$ in the feedback collection and feedback words are assumed to have been drawn from two models: (a) a background model, $p\left( t\left| C \right. \right)$ and (b) a topic model, $p\left( t\left| \theta _F \right. \right)$. Maximizing this log-likelihood using expectation-maximization (EM) algorithm, $\theta _F$ will be calculated. After calculating a language model for feedback questions, we integrate it with the language model from ELMo-based similar questions and the initial language model of the input question (\textbf{expELMoPRF}) to form the expanded input question's LM:
\begin{equation}
\label{equ:RM3_ELMo_feedback}
p\left( t\left| \hat{Q}^{in} \right. \right) =\left( 1-\alpha-\beta \right) p\left( t\left| Q^{in} \right. \right) +\alpha p\left( t\left| \theta_{F_{ELMo}} \right. \right)+\beta p\left( t\left| \theta_{F} \right. \right)
\end{equation}

\subsection{Selective Expansion of questions}
In all the previous question expansion methods, we treat all words appearing in the question as being of the same importance in conveying the main intention of the question while in fact some words play a more central role.
\begin{table}[h]
	\normalsize
	\caption{Importance of different terms in a question}
	\centering
    \begin{tabular}{c|c}
        Q1 & Can anyone suggest a quiet restaurant in Manchester? \\ \hline
        Q2 & Looking for a coffee place in Munich, I'm new.
    \end{tabular}
    \label{table:whycentralitybased}
\end{table}
Take for example the first question in table \ref{table:whycentralitybased} where the asker is seeking a \textit{restaurant} in \textit{Manchester}. Suggesting a \textit{coffee place} in \textit{Manchester} to the asker of $Q1$ instead of a \textit{restaurant} might still be considered relevant to his information need. However, by expanding the word, \textit{Manchester}, other city names which are of the most semantic closeness to it would be added to $Q1$ and a question like $Q2$ that is looking for a \textit{coffee place} in another city, although is unfavourable, would be suggested to the asker as a relevant question.
\par In order to tackle the aforementioned problem, we need to exclude the words that have a more key role in the input question from the expansion process. To this end, we utilized a method proposed by paik et al. \cite{paik2014fixed} to find important the words in questions,. This method sorts the words of a query according to their importance (or so called centrality) in the question taking into account the words frequency in the feedback docs' collection.
Due to space limitation, we omit the full details of the algorithm that calculates the centrality of each word. for each question $Q^{in}$ a vector, $A$, of size $|Q^{in}|$ is calculated in which an element of index $i$ stores the centrality of word $i$ in the input question. This vector is then multiplied by a function of $IDF(t)$ as follows that marks the overall importance of term $t$ in the collection and the contribution of IDF in this function is regularized by a free parameter $c$:
\begin{equation}
\label{equ:didf}
didf\left( t \right) =\frac{idf\left( t \right)}{c+idf\left( t \right)}
\end{equation}
\begin{equation}
\label{equ:final_At}
I\left( q_i \right) =A\left( q_i \right) \cdot didf\left( q_i \right)
\end{equation}
$I\left( q_i \right)$ is the final centrality for each term in the input questions. We choose the word with the most centrality in each question as its central word and omit this word from expansion. We also observed that the central word in some input questions changes before and after applying the effect of IDF and also in those cases, the two words together incorporate the main intention of the question. some of these cases are shown in table \ref{table:central words} in which blue denotes the central word before applying IDF and red denotes the central word after it.
\begin{table}[h]
	\caption{Central words in different questions}
	\centering
    \begin{tabular}{c|c}
\multicolumn{1}{|c|}{Question ID } & \multicolumn{1}{c|}{Question Title} \\ \hline
$Q^{123}$& How do i {\color{blue} fix} my {\color{red}{camcord}}?                     \\
$Q^{177}$& {\color{blue} {Nintendo}} {\color{red} {ds}} lite not charging or turning on                     \\
$Q^{1302}$& How do i {\color{blue} {play}} music with my {\color{red} {keyboard}}?                      \\
$Q^{215}$& How can i have {\color{blue} {laptop}} and tv with the same  {\color{red}{display}}                      \\
$Q^{100}$& What are the best brands of {\color{red} {cheerleading}} {\color{blue} {Shoe}}                     
    \end{tabular}
    \label{table:central words}
\end{table}
For these cases we take both of the words before and after applying IDF to be the central word set for the question. As a result some questions will have one and some others have two central words, and in this work the set of central words for an input question is shown with $C^{Q^{in}}$.
\par We now proceed with removing central words from the expansion process. In \textbf{expAL} method we simply don't find close words for the central word(s) of the input question (\textbf{expAL-centrality}). In \textbf{expKuzi} we exclude the central word(s)' vector(s) in calculating the input question's representation (\textbf{expKuzi-centrality}). Consequently, the euq. \ref{equ:sarkuziCentcalc} becomes:
\begin{equation}
	\label{sarkuziCentcalc_excludeCents}
	\overrightarrow{{Q}_{Cent}^{in}}  = \sum_{q_i\in Q^{in}, q_i \notin C^{Q^{in}}} \overrightarrow{{q}_{i}}
\end{equation}
In Elmo based expansion, \textbf{expELMo}, after calculation contextualized representation vectors for questions' words, we omit the vector representation calculated for the central word  in both the input and the candidate questions as a result of feeding them to the ELMo model, and after that calculate vector representations for those questions not considering their central word. Eventually, we Find the $k$ semantically closest questions to the input question using these new representations. For questions with two central words, this process is repeated with the second word as well and another set of size $k$ is extracted. The final set of questions used for expansion will lie at the intersection of these two sets calculated for each central word (\textbf{expELMoPRF-centrality}). More formally, the equations \ref{equ:ELMo_Vector_rep_in} and \ref{equ:ELMo_Vector_rep_candid} will become:
\begin{equation}
\label{equ:Elmo_centralities removed}
	\forall t_i\in C^{Q^{in}} :
\end{equation}
$$
\overrightarrow{{V}_{ELMo}^{in}}  = \sum_{q_i\in Q^{in},q_i\ne t_i} \overrightarrow{{q}_{i}}
$$
$$
\overrightarrow{{V}_{ELMo}^{C}}  = \sum_{q_j\in Q^{C},q_j\ne t_i} \overrightarrow{{q}_{j}}
$$

\section{Experiments}
In this section we explain the conducted experiments to test the performance of the proposed question expansion methods in question retrieval.
\par
We used two data sets to perform our experiments. The first one is Yahoo! L6 data set \footnote{\url{https://webscope.sandbox.yahoo.com}} that consists of 4,483,032 questions with their best answers, answers and other metadata such as the asker's id. The second data set that we used for evaluation is created by Zhang et al. \cite{zhang2016learning} in which they crawled 1 million questions (a.k.a. question titles)
associated with descriptions (a.k.a. question bodies) and answers from Yahoo! Answers\footnote{This data set is available at \url{https://github.com/ComputerHobbyist/cqa}}. From this crawled questions' set, they have chosen 1260 questions as input questions for each of which there are on average 15 questions retrieved and labeled relevant or non-relevant to their input question which collectively make the entire candidate questions collection. A summary of the statistics for this data set can be found in table \ref{table:Test_data_stats}.
\par All baselines have been implemented using Lemur toolkit\footnote{\url{http://www.lemurproject.org}}. To implement the rest of the methods, however, including the proposed methods, we utilized python 3. 

\begin{table}
	\caption{Test data statistics}
	\centering
\begin{tabular}{|c|c|}
\hline
Number of crawled questions & 1,076,425 \\ \hline
Total number of crawled answers & 7,567,268 \\ \hline
Number of unique words in the collection & 279,353 \\ \hline
Number of test input questions & 1,260 \\ \hline
Number of test candidate questions & 23,917 \\ \hline
Average question length (in words) & 7.5 \\ \hline
\end{tabular}
\label{table:Test_data_stats}
\end{table}
\subsection{Baselines}
We use four methods as our baselines. Our first baseline in question retrieval is the \textbf{BM25} scoring function. \cite{robertson1977probability}. The second baseline is the language modeling approach where we use the \textit{KL-divergence score} with \textit{Dirichlet-Prior smoothing} between the input question's language model and that of a candidate question (\textbf{LMIR}). The third baseline is the \textit{Translation-based language model} proposed by xue et al. \cite{xue2008retrieval} which extends the query likelihood formula taking into account the translation probability between query and document words while calculating the relevance score for a document (\textbf{TR-LM}). The intuition behind this model is that the scoring formula in this model increments the score of a document if there are words in it that are probable to be translated into a query word which itself might have not occurred in the document. For this baseline, in order to extract the translation probabilities between words, we used the mutual information-based method proposed by karimzadegan et al. \cite{karimzadehgan2010estimation} and Yahoo! L6 data set was used for this purpose. The fourth baseline that we compare our proposed methods with is the pseudo relevance feedback model. In this model, we use the generative mixture model \cite{zhai2001model} to estimate the language model of the feedback questions (\textbf{LM-PRF}).
\subsection{Parameter Tuning}
In order to evaluate the proposed methods, we should have come up with the best parameters setting for each method. Consequently, we broke the test data set into two equally sized data sets in terms of test input questions. One is used for parameter tuning (the development set), and another for testing the proposed methods (the testing set) using the best value for each method's parameters resulted from testing them on the development set. For the \textit{LM-PRF} baseline method, experiments on the development set for the best number of feedback questions to use for expansion were carried out and using the first 2 retrieved questions as the feedback collection yielded the best result in terms of MAP. Following this observation, for all the experiments in which whether PRF is interpolated with ELMo or used alone, we use the first two retrieved questions as the feedback set. Similarly, the best number of semantically similar questions to the input question, $k$, to be used in ELMo-based expansion methods was chosen from the set $\left[1,10\right]$ and the best result was obtained while $k=5$.
\par
In the \textit{expAL} method, $\alpha$ values for the weights of the expansion words in period $\left[0.2,0.4\right]$ were tested and 0.4 was set. As for the number of expansion words per each base word, we tested 1 expansion word per base word to 10 expansion words and settled for 2 expansion words per each base word. In the \textit{expKuzi} method, we explored from value 8 for the number of expansion words, $v$, for the entire question to value 20 and the best result was achieved when $v=9$.
\par Interpolating the expansion LMs with the initial input question's LM, we set $\lambda = 0.65$ in \textit{expKuzi} and \textit{expKuzi-centrality}. In \textit{expELMo} $\alpha$ was set to 0.3 and in \textit{expELMoPRF-centrality}, $\alpha$ and $\beta$ were set to 0.3 and 0.2 respectively.
\par In the experiments with selective expansion of question words where central words calculation is needed, we used 10 pseudo relevance feedback questions to take into account the importance of input question terms in those questions. The number of iterations for the EM algorithm to calculate the centrality of each word in the question was set to 12 as it was recommended by the authors of that method.
\subsection{Evaluating Results}
\par Table \ref{tabel:results_non_selective} presents the performance of the baseline methods in retrieving the relevant questions for an input question in terms of Mean Average Precision (MAP), along with the performance of the proposed methods in this retrieval task. In spite of the fact that \textit{LM-PRF} is a strong baseline and \textit{expAL} and \textit{expKuzi} which expand based on semantic similarity at word level have not been able to outperform it on their own, our results proved that using query expansion methods can be useful in dwindling the lexical gap that exists between an input question and the question that might be relevant to it. Another interesting point is that since \textit{expELMo} have been able to achieve higher performance compared to \textit{LM-PRF}, we can conclude that using questions that have been identified to carry the same intention as the input question by \textit{expELMo} method, are more accurate compared to the questions that are suggested by \textit{LM-PRF}. The reason for this claim is that although both methods work in similar ways, since the questions extracted as feedback for \textit{expELMo} consider not only the lexical similarity at some level, but also both the semantic and syntactic similarity between questions as well as a result of deep word representations, \textit{expELMo} is able to extract better feedback questions for expansion.

\begin{table}[ht]
	\caption{Performance comparison of baselines and proposed question expansion methods in terms of MAP in the question retrieval system}
	\centering
\begin{tabular}{c|c|c|}
\cline{2-3}
\multicolumn{1}{l|}{}                                                                             & Method Name & MAP    \\ \hline
\multicolumn{1}{|c|}{\multirow{4}{*}{Baselines}}                                                  & BM25        & 0.7267 \\ \cline{2-3} 
\multicolumn{1}{|c|}{}                                                                            & LMIR        & 0.7280 \\ \cline{2-3} 
\multicolumn{1}{|c|}{}                                                                            & TR-LM       & 0.7291 \\ \cline{2-3} 
\multicolumn{1}{|c|}{}                                                                            & LM-PRF      & 0.7328 \\ \hline
\multicolumn{1}{|c|}{\multirow{3}{*}{\begin{tabular}[c]{@{}c@{}}Proposed\\ Methods\end{tabular}}} & expAL       & 0.7315 \\ \cline{2-3} 
\multicolumn{1}{|c|}{}                                                                            & expKuzi     & 0.7316 \\ \cline{2-3} 
\multicolumn{1}{|c|}{}                                                                            & expELMo     & 0.7375   \\ \hline
\end{tabular}
\label{tabel:results_non_selective}
\end{table}

\par In table \ref{table:results_selective_expansion} the results of the proposed methods considering the selective expansion have been given. Improvements of the proposed methods in this table have all been statistically significant over the baseline methods in terms of MAP using t-test with 95\% confidence level. Our last proposed method which is the ELMo-PRF based question expansion, has gained the most significant result. In addition to all baseline methods, its improvement over the two word-similarity based methods, \textit{expAL-centrality} and \textit{expKuzi-centrality}, has been significant as well. The ELMo-PRF based question expansion method has been able to improve the question retrieval performance by $1.1\%$ and $1.8\%$ over \textit{expAL-centrality} and \textit{TR-LM} respectively.

\begin{table}
	\caption{Performance of the proposed methods when selective expansion has been applied on them. * Signals statistically significant improvement over \textit{BM25, LMIR} and \textit{TR-LM} using t-test, and \#, denotes statistically significant improvements over \textit{LM-PRF, expAL-centrality} and \textit{expKuzi-centrality}}
	\centering
\begin{tabular}{|c|c|}
\hline
Method Name            & MAP    \\ \hline
expKuzi-centrality       & $0.7324^{*}$ \\ \hline
expAL-centrality     & $0.7342^{*}$ \\ \hline
expELMo-centrality     & $0.7384^{*}$ \\ \hline
exp-ELMoPRF-centrality & $0.7428^{*\#}$ \\ \hline
\end{tabular}
\label{table:results_selective_expansion}
\end{table}

\section{Conclusion and Future Work}
In this work we explored question expansion to enhance the question retrieval system's performance in CQA systems. We utilized two already proposed word embedding-based expansion methods in addition to our proposed ELMo-based question expansion method. we further improved these methods by proposing not to consider all words in the expansion process by finding the most central words and removing them while expanding. Our experiments showed that question expansion has been successful in reducing the lexical gap with the ELMO-PRF based expansion method as being the most successful. In addition, we showed that avoiding the expansion of some words which are key words, can be beneficial as by not changing a question's main intention, the chance of finding more relevant questions to it increases and this action actually brought improvements to the proposed methods. We also showed that using ELMo-based similar questions as feedback docs outperforms using questions elicited by PRF. For future, other methods of finding the most central words with higher accuracy could be incorporated in the expansion process. Considering the fact that using ELMo representations that encode syntactic and semantic features at sentence level has proven efficient, newer methods for word and question embedding could also be investigated in future works both for query and question expansion.

\printbibliography
\end{document}